# Efficient terahertz optical filtering with large-area all-metal and polymer-metal woven wire meshes

SIMON ROSSEL, WENTAO ZHANG, HASSAN A. HAFEZ, SAVIO FABRETTI, AND DMITRY TURCHINOVICH*

*Fakultät für Physik, Universität Bielefeld, Universitätsstraße 25, Bielefeld 33615, Germany*
*dmtu@physik.uni-bielefeld.de*

**Abstract:** Many components for terahertz (THz) optical filtering are mechanically fragile and are hard to produce with large aperture, making them unsuitable for applications where larger THz beam diameter is required. In this work, the THz optical properties of industrial-grade, readily available and inexpensive woven wire meshes are studied using THz time-domain spectroscopy and numerical simulations. These meshes are meter-sized, free-standing sheet materials that are principally attractive for the use as robust, large-area THz components. Our results show that such meshes can act as efficient, tunable THz bandpass filters due to sharp plasmonic resonance supported by the interwoven metallic wires. Further, the meshes that combine metallic and polymer wires act as efficient THz linear polarizers with a polarization extinction ratio (field) above 60:1 for frequencies below 3 THz.

## 1. Introduction

The terahertz (THz) frequency range offers great potential for many technical applications, e.g. high-speed wireless communication, security screening, imaging and sensing, and is widely used for fundamental research on materials [1-5]. With the growing importance of the THz applications, efficient control of THz signals, such as frequency and polarization filtering, is required. THz filters and linear polarizers are fundamental optical components, which are integral part of advanced THz technology. THz polarization filtering capabilities of various structures have been demonstrated, e.g. parallel metallic bars [6-8], aligned carbon nanotubes [9] and liquid crystals [10], and a large variety of metamaterials for THz frequency filtering [11-19] have been reported. There are also Bragg mirrors for THz [20]. These structures are either free-standing, or grown on a substrate e.g. using photolithography, nanoimprinting [7] or inkjet printing [8]. Wet etching and laser micromachining [15] can be used to create the free-standing structures. The complexity and cost of some of the manufacturing processes involved make it difficult to produce large aperture filter devices. Furthermore, the substrate of the substrate-based structures can induce unwanted etalon effects, whereas the free-standing counterparts can be mechanically fragile. The commercially available wire grid polarizers made of thin conductive wires and frequency filtering metamaterials are usually rather costly. Hence, a large-area, free-standing, robust and low-cost structure with small feature size for THz optical filtering is still needed.

Woven wire mesh fulfils these conditions, as it is an industrial-grade, self-supporting sheet material produced in meter dimensions, while possessing a complex micrometer-sized pore geometry. Such meshes are typically used for filtering of particles, liquids and gasses, mechanical reinforcement in products, electromagnetic shielding etc. Woven wire mesh is woven similar to cloth with two distinct, perpendicular and repeatedly crossing wire directions, which form a periodic wire structure with air gaps. A great advantage of woven wire meshes is the possibility to control THz signals with beam waists up to a meter. Such THz beams are encountered in, for example, THz wireless communication, THz space applications, and THz imaging [1-3,21]. It has been shown that metal wire structures can be used for various purposes, e.g. as negative refractive index metamaterial [22]. There is a large number of woven wire mesh types, which can be categorized by their material composition, such as all-metal and polymer-metal. The all-metal plain weave type, which has a square-hole-array-like woven metallic geometry, has recently drawn the attention of the THz community due to its interesting THz optical properties [23-26]. All-metal plain weave has shown frequency bands with up to 88% transmittance as well as abnormal group velocities [23], sharp plasmonic resonances, which are induced by the bent shape of the woven metal wires [24,25], and mode splitting phenomena when the THz is at oblique incidence [26]. Sharp plasmonic resonances can make the all-metal plain weave attractive not only for sensing applications but also for THz frequency filtering. However, favorable structural parameters and configurations for the realization of THz frequency filters using plain weave have not been further investigated. Meshes combining metallic and polymer wires have also been studied in the THz and GHz frequency range [27-29]. Their polarization-dependent electromagnetic response [29] makes polymer-metal mesh interesting for polarization filtering.

In this work, the THz optical properties of all-metal plain weave and a polymer-metal mesh are investigated using THz time-domain spectroscopy (THz-TDS) and numerical simulation, and the THz filtering performance of the meshes is demonstrated. The experimental and numerical results show that the THz optical properties of the plain weave are dominated by plasmonic resonances. A passband with sharp band edges is formed between two adjacent resonances, making the plain weave a low-cost, readily available bandpass filter. In turn, the polymer-metal mesh acts as an efficient THz linear polarizer with high polarization extinction ratio.

## 2. Methods and materials

Woven wire meshes are periodic structures formed by two perpendicular wire directions, called warp and weft. These run alternately above and below the wires of the respective other wire direction. In the case of plain weave, the wires form square holes in between them, as shown in Fig 1(a). The unit cell of the plain weave is composed of four square holes with a period $\Lambda$, as illustrated in Fig. 1(b). The woven geometry of the wires is modelled using bent sections at the wire meeting points and straight segments connecting the former. The structural parameters to describe the shape of the wires are the wire diameter $d$, the bending heights $h_x$, $h_y$ and bending radii $R_x$, $R_y$ of the warp wires in the *x*-direction and weft wires in the *y*-direction. Similar geometrical considerations can be applied to the polymer-metal mesh as well. Two plain weaves, a stainless steel one and a bronze one, are investigated. The bronze plain weave and the polymer-metal mesh were supplied by the company GKD – Gebr. Kufferath AG. A commercially available THz-TDS setup (Teraflash, Toptica Photonics AG) was used to measure the transmission of collimated, linearly polarized THz radiation through the woven wire meshes. The diameter of the collimated beam was 5 cm. In later experiments, the plain weaves are rotated in the collimated beam so that the THz radiation is at oblique incidence.

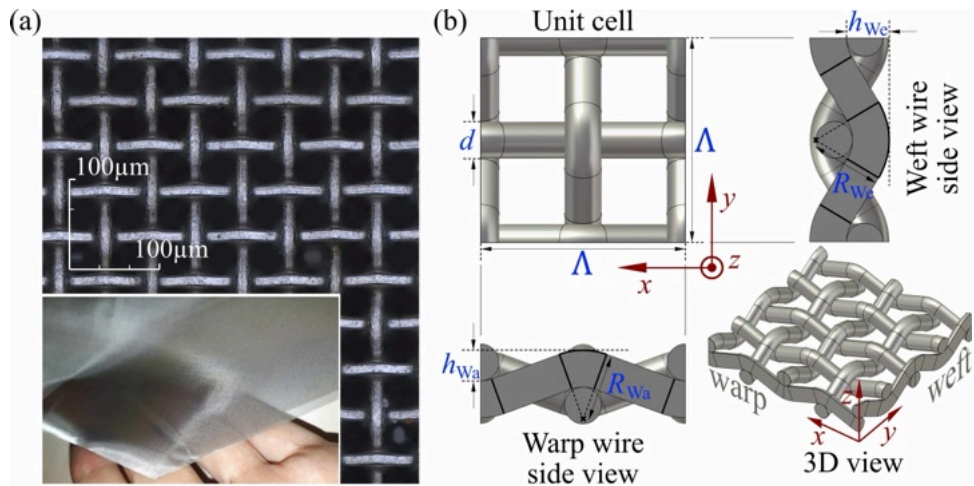

Fig. 1. (a) Microscopic image and photograph of stainless steel plain weave, and (b) the simulation model.

In addition to the experiment, the transmission through the modelled unit cell of the plain weave is simulated using the finite element method - based frequency domain solver of the commercial software CST Studio Suite. In the simulation, the metallic wires are assumed to be perfect electric conductors. Unit cell boundary conditions are applied to the model in the transverse *x*- and *y*-directions and Floquet ports are used in the *z*-direction. The calculation domain is excited by plane waves with different angles of incidence using the fundamental transverse electric and transverse magnetic Floquet modes. It is noted that the modelled unit cell assumes a smooth, circular cross-section of the wires and therefore does not take into account any deformation of the wire shape due to mechanical stress. The contact points of the orthogonal wires of the simulation model are perfectly conducting connections, while in the experiment there is a contact resistance. In the simulation, additional small PEC spheres also had to be introduced at the wire meeting points, each occupying the submicron volume in between the approaching wires (sphere diameter is half the diameter of the wires *d*). Furthermore, the industrial-grade mesh shows a degree of imperfection, making it different from the idealized simulated geometry. These discrepancies can in the following lead to certain differences between experimental and simulation results.

## 3. Results and discussion

*3.1 Plasmonic resonances induced by woven geometry of all-metal plain weave*

The first investigated sample of stainless steel plain weave has, according to manufacturer information and the microscopic images, a period of $\Lambda = 100\,\mu\text{m}$, a wire diameter of $d = 18\,\mu\text{m}$ and warp and weft wire bending heights of $h_x = 15\,\mu\text{m}$ and $h_y = 21\,\mu\text{m}$. The bending radii $R_x = 33.75\,\mu\text{m}$ and $R_y = 36\,\mu\text{m}$ used in the modelling were determined through the comparison of experimental and simulation results, which is presented later on. The warp wires of the plain weave are overall straighter than the weft wires. However, the bent sections of the warp wires are bent more with a smaller bending radius compared to the bent sections of the weft wires. This difference in the geometry of the warp and weft wires is due to the treatment in the manufacturing process. The measured field transmission of the THz radiation at normal incidence on the plain weave is shown in Fig. 2(a) for the THz electric field polarization along the *x*- and *y*-direction by the blue and red symbols, respectively. The transmission through the subwavelength holes increases with frequency, as the square holes can be regarded as rectangular cutoff waveguides. Two resonances are measured with transmission dips at 3.23 THz for the blue spectrum and 3.40 THz for the red spectrum, roughly corresponding to dip wavelengths of $\lambda \sim 0.93 \times \Lambda$ and $0.88 \times \Lambda$, respectively. The resonance of the blue spectrum is much stronger than the one of the red spectrum and shows the asymmetric Fano shape. The simulated field transmission spectra, depicted for the THz electric field polarization along the *x*- and *y*-direction by the dark red and dark blue lines, respectively, show that the simulated low-frequency transmission and Fano resonances are in good agreement with the corresponding experimental spectra. However, the strength of the measured resonances is weaker due to the differences between the experimental sample and simulation model. From now on, these two Fano resonances are referred to as Fano resonance type A (F-A). Previous study of plain weave [24] identified F-A as Wood anomaly, which is, according to theory [30-33], related to the excitation of surface plasmons. Wood anomalies have also been observed in other metallic structures at THz frequencies [34-37]. The simulated electric field distributions at dip frequency of the F-A depicted in Figs. 2(b)-(c) and Figs. 2(d)-(e) for the polarization along the *y*- and *x*-direction, respectively, show localized electric fields on top of the straight wire segments of the wire direction in the respective polarization direction. The *y*-component of the electric field has opposite signs at the back and front face of the wires, as shown by the inset in Fig. 2(a) for incident *y*-polarization. When the incident THz radiation is polarized in the *x*-direction, the same can be found for the *x*-component of the electric field. These fields resemble the fields of the surface plasmons excited on the wires the in polarization direction. For the polarization along the strongly bent weft wires in the *y*-direction, the dip is at lower frequencies compared to the polarization along the warp wires in the *x*-direction. In contrast, the simulated peak frequencies of F-A for the two polarization directions differ by only 32 GHz. The measured F-A with the polarization along the warp wires in the *x*-direction is particularly weak, which could be caused by a more inconsistent warp wire shape of the experimental sample.

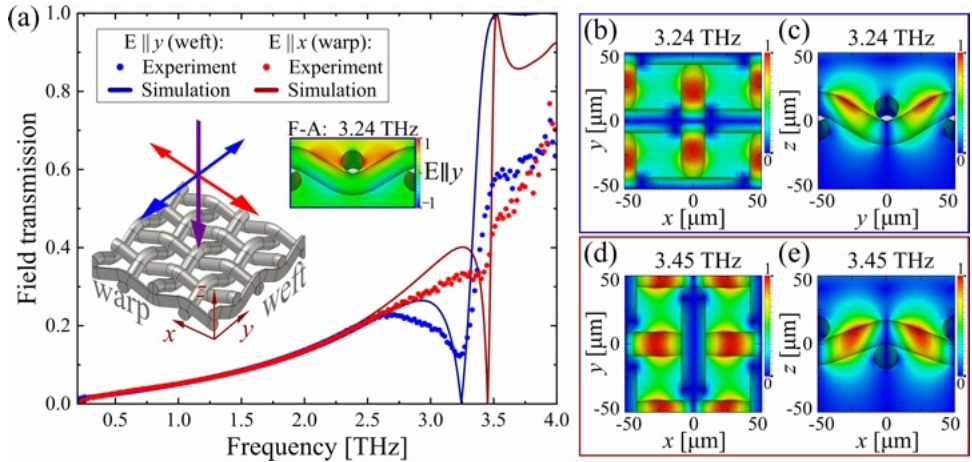

Fig. 2. (a) Measured and simulated transmission spectra of stainless steel plain weave. Normalized electric field distributions of the front face and side view of the wires in polarization direction at dip frequencies for (b), (c) *y*-polarization and (d), (e) *x*-polarization. Field values are displayed on the respective cut plane and on the surface of the wires that intersect the planes. The inset in (a) shows the normalized distribution of the *y*-component of the electric field at dip frequency for incident *y*-polarization.

    Next, the zero-order transmission through the plain weave at oblique incidence of *p*-polarized THz radiation with the plane of incidence along each of the two wire directions is investigated. First, the results for the plane of incidence along the weft wires (*yz*-plane) are presented. All the measured field transmission spectra of the THz radiation at oblique incidence are condensed into a two-dimensional contour plot shown in Fig. 3(a). The positive and negative angles correspond respectively to clockwise and anti-clockwise rotation of the mesh in the collimated beam. The contour plot shows that the transmission through the plain

weave is independent of the rotation direction, which is due to the symmetry of the mesh. At oblique incidence, F-A is seen splitting into two resonances with different frequencies: one shifting to around 3.73 THz at $\theta=10°$ and the other to 2.39 THz at $\theta=60°$, and similar for negative $\theta$. This splitting into a higher-frequency and lower-frequency resonance is possibly due to the in-plane component of the wave vector of the THz radiation at oblique incidence. This in-plane component can lift the degeneracy of the surface plasmons propagating in different directions on the wires [36]. There are additional resonant transmission features at the spectral positions where a new diffraction order appears. The spectral positions are shown by the dashed lines for the corresponding order $m$ and $n$, denoted as $(m,n)$. These are approximated by $f_{m,n} \cong c/\lambda_{m,n}$, where $c$ is the speed of light and $\lambda_{m,n}$ is the Rayleigh wavelength, which is given, according to grating theory [32], by

$$\lambda_{m,n} = \frac{-B+\sqrt{B^2+A\cos^2(\theta)}}{A}, \qquad (1)$$

where $A=(m^2+n^2)/\Lambda^2$, $B=(m\cos(\varphi)+n\sin(\varphi))\times\sin(\theta)/\Lambda$ and $\varphi$ is the angle of the plane of incidence with the x-axis. These resonant features are referred to as Rayleigh-Wood anomalies (RWA) [30-32]. The differences of the measured and calculated spectral positions of the RWA orders are most likely due to slight, unavoidable misalignments of the angle of incidence $\theta$ in the experiment.

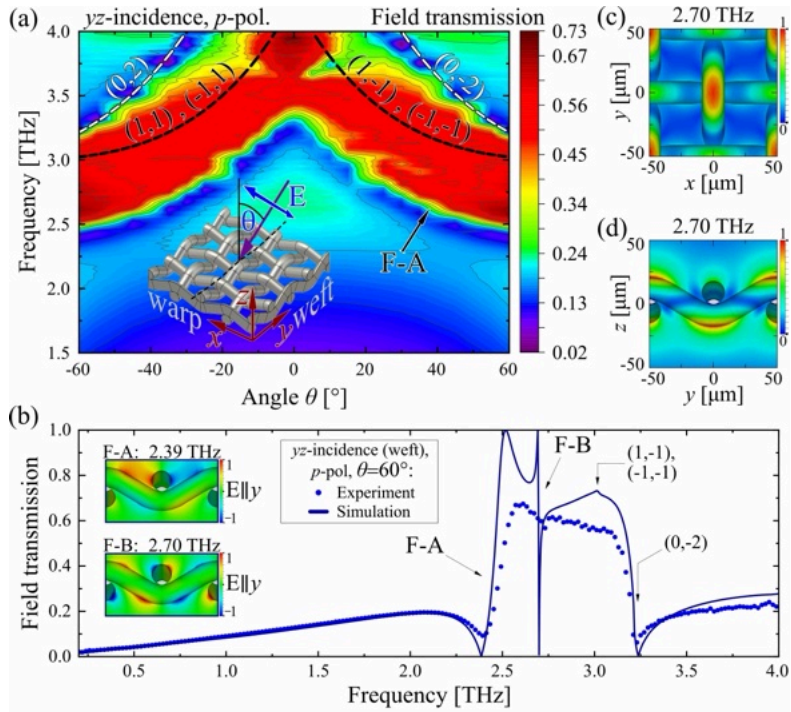

Fig. 3. (a) Contour of measured field transmission spectra with yz-plane of incidence, (b) the measured and simulated spectra at $\theta=60°$ incidence and (c), (d) the electric field distributions at 2.70 THz for $\theta=60°$ incidence. The insets in (b) show the normalized distributions of the y-component of the electric field at dip frequency of F-A and F-B.

In Fig. 3(b) the measured field transmission spectrum at $\theta = 60°$ incidence is shown together with the simulated one, by the blue symbols and dark blue line, respectively. Besides F-A and the two RWA orders, a second Fano resonance of inverted symmetry is simulated with 2.69 THz peak frequency and 2.70 THz dip frequency. This Fano resonance is referred to as Fano resonance type B (F-B) from hereon. The measured F-A and F-B are weaker than those in the simulation due to the previously described discrepancies between the experimental sample and the model. Further, the 22 GHz frequency resolution of the experiment cannot fully resolve the narrow profile of F-B. Even though F-B is not discernible in the contour plot, a shift of the resonance with the angle of incidence $\theta$ is indicated in the individual spectra, similarly as shown in preceding results for xz-incidence. The simulated electric field distributions at dip frequency of F-B, depicted in Figs. 3(c)-(d), show localized fields at the bent weft wire sections. The two insets in Fig. 3(b) show the distributions of the y-component of the electric field at dip frequency of F-A and F-B. Here, the fields on the weft wires in the two halves of the unit cell divided along the vertical have opposite signs. The difference between the field distributions of F-A and F-B is that the fields of F-A are also inverted at the back and front faces of the weft wires. Most importantly, a transmission band with up to 67% field transmission is measured between the transmission dip of F-A at 2.39 THz and the (0,-2) RWA order at 3.23 THz. This makes this constellation interesting for use as a THz bandpass filter with the advantage that the band edges are two plasmonic resonances whose frequency can be tuned by rotation of the mesh.

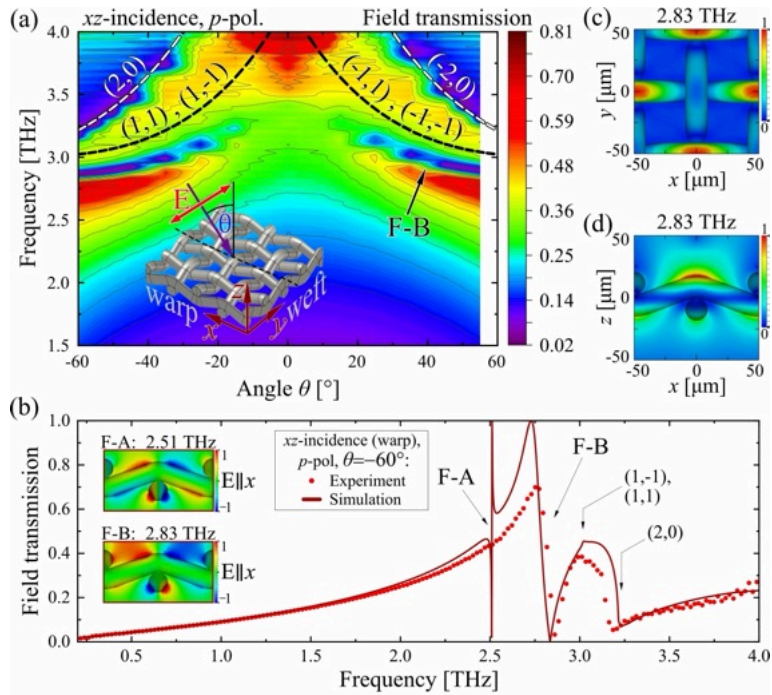

Fig. 4. (a) Contour of measured field transmission spectra with xz-plane of incidence, (b) the measured and simulated spectra at $\theta = -60°$ incidence and (c), (d) the electric field distributions at 2.86 THz for $\theta = -60°$ incidence. The insets in (b) show the distributions of the x-component.

The results of the THz radiation incident on the plain weave with the plane of incidence along the warp wires (xz-plane) is presented in Figs. 4(a)-(d). The measured field transmission contour plot, shown in Fig. 4(a), differs significantly from the previously presented results. F-A is not discernible in the contour plot. The strong Fano resonances, whose dip shifts from 2.86 THz at $\theta = -60°$ to 3.75 THz at $\theta = 10°$ and similar at positive angles, is attributed to F-B. Therefore, F-B splits, similar to F-A, into a higher- and lower-frequency resonance at oblique incidence. Its increase in resonance strength with the angle of incidence $\theta$ indicates a strong interaction with the z-component of the electric field. The measured field transmission spectrum at $\theta = -60°$ incidence shown in Fig. 4(b) shows a large field transmission contrast between the peak of F-B of 70% and its dip of 3%, which are only separated by 108 GHz. Such a strong and sharp resonant transmission phenomenon could possibly be used for efficient THz sensing applications [38]. The simulated electric field distributions at dip frequency of the F-B, depicted in Figs. 4(c)-(d), show localized fields at the bent wire sections of the warp wires. This localization at the wires along the plane of incidence is the same as for F-B for yz-incidence (Figs. 3(c)-(d)), and the distributions of the x-component at the dip frequency of the resonances, shown by the insets in Fig. 4(b), are similar to the distributions of the y-component in Fig. 3(b). The transmission dips at 3.03 THz and 3.24 THz in Fig. 4(b) correspond to RWA.

*3.2 Plasmonic resonance dependence on structural parameters*

Fig. 5(a) shows the simulated field transmission spectra for plain weave with $d = 18\,\mu m$, $h_x = h_y = d$, $R_x = R_y = 36\,\mu m$ and varying periods $\Lambda$ at $\theta = 60°$ incidence of the p-polarized THz. For this symmetric plain weave, it is sufficient to simulate the situation with the plane of incidence along one of the two wire directions. Here, yz-orientation of the plane of incidence is simulated. The spectra show the two Fano resonances and the RWA orders, each located at lower frequency for increasing $\Lambda$. For the frequencies away from the resonances, the transmission grows with increase in $\Lambda$ due to the increasing hole width between the wires. The symbols in Fig. 5(b) show the simulated wavelengths of the RWA orders and the Fano dips. On one hand, the dip wavelengths of F-A and F-B respectively match the linear equations $\lambda_{\text{F-A}}^{60°} \sim 1.25 \times \Lambda$ and $\lambda_{\text{F-B}}^{60°} \sim 0.91 \times \Lambda$, shown by the lines. These Fano resonances can thus be critically controlled by the period $\Lambda$ as well as the angle of incidence $\theta$. On the other hand, the wavelengths of the two RWA orders agree with the dashed lines, which show their expected dependence according to Eq. (1). The spectral feature in Fig. 5(a) at 3.5 THz for $\Lambda = 150\,\mu\text{m}$ is possibly the higher-frequency resonance that has split from either Fano resonance.

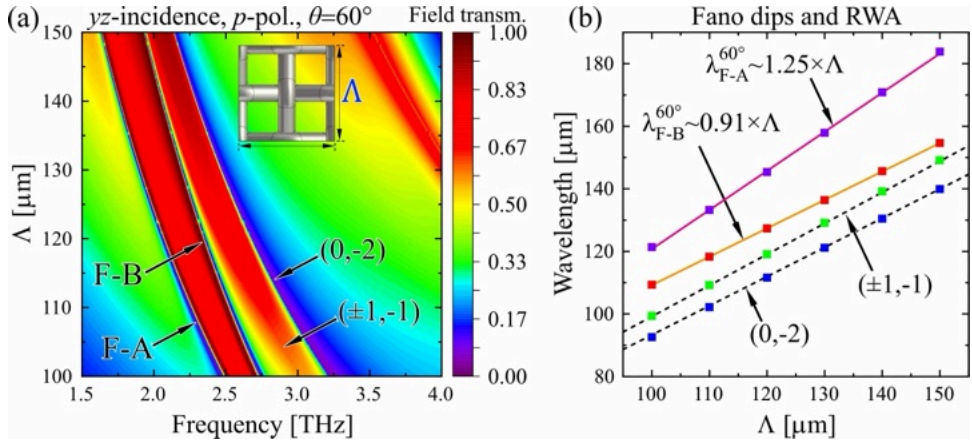

Fig. 5. (a) Contour of simulated transmission through the symmetric plain weave for different periods $\Lambda$ at $\theta = 60°$ incidence and (b) the corresponding wavelengths of the simulated Fano dips and RWA shown by the symbols. The through lines are the fits of the Fano dip wavelengths and the dashed lines show the expected positions of the RWA according to Eq. (1).

Now, the field transmission of the symmetric plain weave with $\Lambda = 100\,\mu\text{m}$ and varying wire diameter $d$ is studied in Fig. 6(a). It is noteworthy that as the wire diameter $d$ of the model increases, the bending heights $h_x$, $h_y$ increase as well and the opening width between the wires decreases. The decrease of the opening width leads to a lower transmission of the nonresonant frequencies, as in earlier simulation. F-A is located at higher frequencies with greater $d$, whereas the F-B shows opposite dependence. The shift of F-A to higher frequencies could be attributed to the increase of the effective plasma frequency of the mesh, similar to that of a three-dimensional periodic metal wire array [23,24,39]. Fig. 6(b) shows the calculated quality (Q) factors of the simulated Fano resonances, defined as $Q = f_r/\Delta f$, where $f_r$ is the resonance frequency and $\Delta f$ the resonance width, which are estimated for the Fano resonances with dip frequency $f_{dip}$ and peak frequency $f_{peak}$ as $f_r = (f_{dip} + f_{peak})/2$ and $\Delta f = |f_{dip} - f_{peak}|$. F-A becomes narrower and its Q-factor increases with smaller wire diameters $d$, reaching a peak Q-factor value of 182 for the smallest simulated wire diameter $d = 8\,\mu\text{m}$. The narrowing of F-A with decreasing bending height is consistent with F-A being induced by the bending of the wires, where F-A vanishes in the limiting case of a metallic hole array with a flat surface [24]. On the other hand, the Q-factor of F-B is lowest at $d = 15\,\mu\text{m}$ and increases for wire diameters above and below.

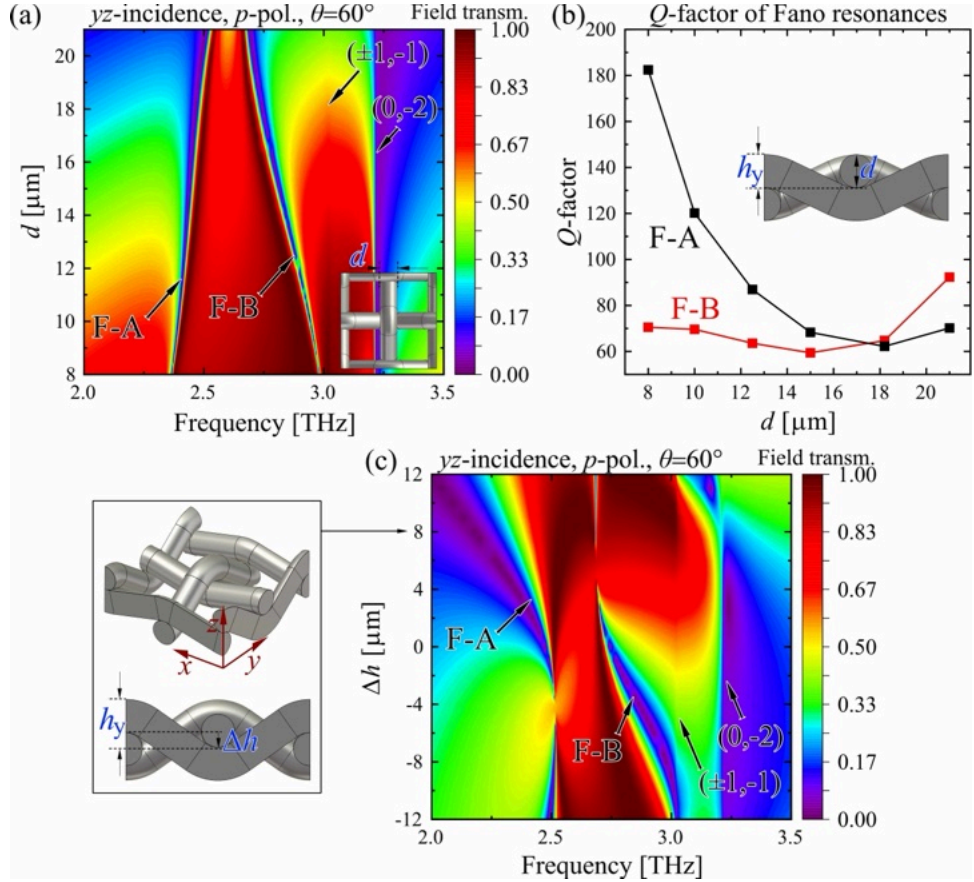

Fig. 6. (a) Contour of simulated transmission of symmetric plain weave for different wire diameters $d$, (b) the calculated Q-factors, and (c) Contour of simulated transmission for varying bending height differences $\Delta h$ of plain weave wire directions.

The influence of the two wire directions having different bending heights relative to one another is briefly discussed in the following. The bending heights of the two wire directions are correlated by $h_x = d - \Delta h$ and $h_y = d + \Delta h$, where $2 \times \Delta h$ is the bending height difference. Fig. 6(c) shows the simulated field transmission for different $\Delta h$. Here, the yz-orientation of the plane of incidence is simulated, which lies along the wires running in the y-direction with bending height $h_y$. The transmission spectra for $|\Delta h| > 4$ μm show sharp resonances around 2.7 THz and 2.5 THz for positive and negative $\Delta h$, respectively, and a broad resonance, which redshifts with $\Delta h$. The behavior over all values of $\Delta h$ is similar to a system of two coupled modes, where a frequency-invariant mode is differently coupled to a $\Delta h$-dependent mode, corresponding to the sharp and broad resonance, respectively. At $\Delta h = 0$ μm an avoided crossing of the modes is observed, and the resonances vanish around $\Delta h = \pm 4$ μm. This vanishing of the resonances at the avoided crossing has been reported for coupled modes in other photonic and plasmonic structures [40,41]. The frequency-invariant mode is frequency shifted due to the coupling. Its frequency is given by the mean value of the frequency of the two sharp resonances at large $\Delta h$ values, which amounts to ca. 2.6 THz. On the other hand, the redshift of the $\Delta h$-dependent mode with increasing bending height of the wires along the plane of incidence is similar to the behavior of the plasmonic mode observed in a single-layer bent wire array [24]. Without considering the different bending radii, the transmission through the stainless steel plain weave in Fig. 3(b), where F-B is narrow, corresponds to $\Delta h = 3\ \mu m$ shown in the contour plot. At the same time Fig. 4(b), where F-A is narrow, corresponds to $\Delta h = -3\ \mu m$. Accordingly, F-A and F-B would correspond to the excitation of the two coupled modes. The decrease in the frequency difference of F-A and F-B with the wire diameter $d$ in Fig. 6(a) could be caused by the decrease in coupling strength of the modes. We emphasize that the wire height difference strongly determines the transmission properties of plain weave. Evidently, the difference of the measured transmission of the stainless steel mesh shown in the previous section (plane of incidence along the warp and weft wires) is due to the difference in bending heights.

### 3.3 THz bandpass filter with sharp band edges of plasmonic resonances

A passband between the dip of F-A and the (0,-2) RWA order can be realized, as shown in Fig. 3(a), when the plane of incidence of the THz radiation is along the strongly bent weft wires. Previous simulation results show that the plain weave should ideally have a small opening width for a large transmission contrast of the passband to the blocked regions and the wire height difference $\Delta h$ should lead to a full suppression of F-B. We note here that the wire height difference $\Delta h$ of a commercial mesh is generally unknown and is not controlled in the manufacturing process.

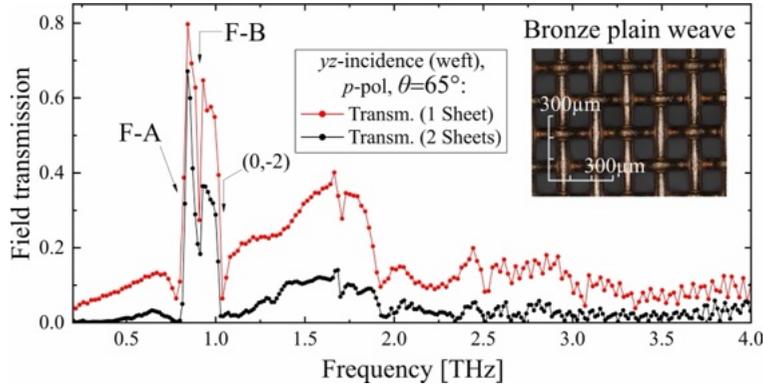

Fig. 7. Measured field transmission of bronze plain weave, where the red and black dots correspond to a single sheet of mesh and a stack of two mesh sheets, respectively.

A bronze plain weave with a wire diameter of $d = 63\ \mu m$ and a period of $\Lambda = 306\ \mu m$ is used to demonstrate the operation as bandpass filter. This plain weave has a small opening width as indicated by its high ratio of $d/\Lambda \sim 0.21$, and the passband will be at lower frequencies compared to the stainless steel mesh due to the larger period. The measured field transmission of the THz radiation at $\theta = 65°$ incidence with the plane of incidence along the weft wires (yz-plane) is shown by the red dots in Fig. 7. Up to 80% transmission is observed between the dip of F-A at 0.80 THz and the (0,-2) RWA at 1.03 THz. Unfortunately, F-B around 0.91 THz is not sufficiently suppressed and disrupts the transmission band. This is due to a too small bending height difference $\Delta h$ of the two wire directions of the bronze plain weave. In principle, a more homogeneous passband could still be achieved by pulling on the warp wires, leading to an increase of $\Delta h$ and a further suppression of F-B. This can be one of the next optimization steps in the further development of THz filters based on the woven wire meshes.

The measured transmission through two stacked up sheets of bronze plain weave is depicted by the black dots, with the second sheet directly behind the other without alignment of their microscopic geometries. The two stacked up plain weaves show a band with up to 60% transmission with a large transmission contrast to the suppressed lower frequencies of the order of 20:1 peak-to-peak and sharp band edges. By changing the angle of the incident THz radiation, the frequency of the passband can be tuned. Generally, the spectral positions of the band edges of the resulting passband can be estimated for any plain weave at different angles of incidence in the following way. Eq. (1) gives the spectral position of the RWA, and the position of the dip of F-A can be estimated by assuming the same relation of the dip frequency to the period at normal incidence of $f \cong c/(0.93 \times \Lambda)$ and relative $\theta$-dependency, as for the case of the stainless steel mesh (Fig. 3(a)). The stainless steel mesh with $\Lambda = 100$ μm described in the previous sections is among the available plain weaves with the smallest structure size, which has the highest frequency passband around 3 THz. By using plain weave with larger periods like the bronze plain weave, it is possible to realize passbands for any desired frequency below that.

*3.4 Polymer-metal mesh THz linear polarizer*

The second mesh type investigated in this work is the polymer-metal mesh, acting as an efficient broadband THz linear polarizer. Its structure, as shown in Figs. 8(a)-(b), combines stainless steel and PET wires. Fig. 8(b) serves here as a CAD illustration of the mesh. For this mesh, no simulation is performed. We note here that previous studies have simulated meshes consisting of polymer and metallic components [27-29]. The metallic warp wires along the *x*-direction with a wire diameter of 90 μm are spaced by 692 μm, while the metallic weft wires along the *y*-direction with a wire diameter of 50 μm have a spacing of only 71 μm. The frequency selective elements are the closely-spaced metallic weft wires held in place by the other wires. Thus, the working principle is analogous to commercial THz wire-grid polarizers, only that the polymer-metal mesh is a self-supporting structure due to the additional wires, which predestines it for the use as large-area component up to the meter dimension. Fig. 9 shows the measured transmission of the THz radiation at normal incidence on the mesh with the polarization in the *x*- and *y*-direction. Here, the transmission of the THz radiation polarized parallel to the closely spaced weft wires in the *y*-direction is significantly more suppressed than that of the perpendicular direction, shown by the red and blue dots, respectively. Several weak resonant transmission phenomena are observed below 1.5 THz, most likely due to similar plasmonic effects as in plain weave. The polarization extinction ratio (PER) for the THz field, represented by the black dots, is given by the ratio of the blue transmission spectrum to the red one and lies above 60:1 at frequencies below 3 THz. This corresponds to the power PER in excess of 3600 for the THz frequencies in the range of ca. 0.2 – 3 THz. Despite the additional absorption and scattering losses due to the wires for mechanical stabilization of the mesh, the observed PER values are comparable to that of commercial free-standing THz wire-grid polarizers [42]. It is worth noting that the incident and transmitted $x$-polarized ps-scale single-cycle THz pulses used in our experiment are almost identical in shape, indicating that the polymer-metal mesh does not exhibit any significant dispersion. Thus, the polymer-metal mesh acts as an efficient and low-cost THz linear polarizer, which is mechanically robust and readily available in meter-sized sheets that do not need special framing.

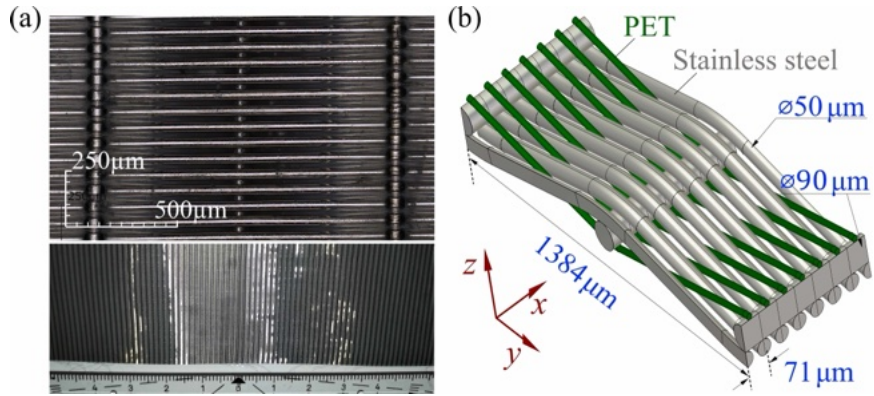

Fig. 8. (a) Microscopic image and photograph of the polymer-metal mesh, and (b) a CAD illustration of several unit cells of the mesh.

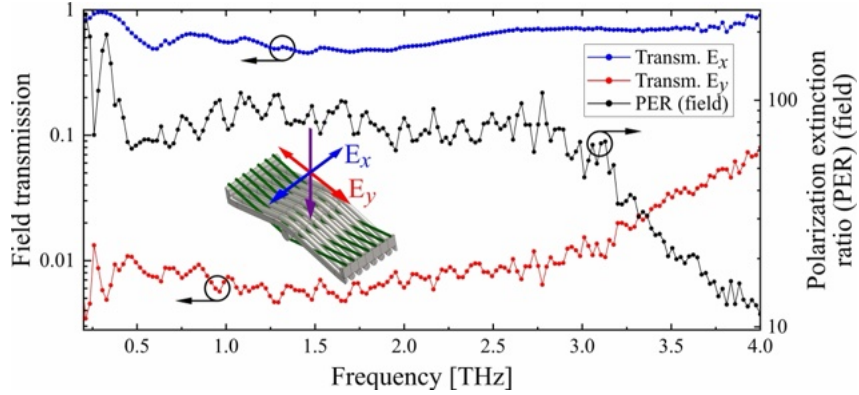

Fig. 9. Measured transmission and calculated polarization extinction ratio for the polymer-metal mesh.

## 4. Conclusion

In this work, efficient terahertz optical filtering with large-area all-metal and polymer-metal woven wire meshes was studied. The plasmonic resonances of all-metal plain weave was experimentally and numerically investigated at THz frequencies for obliquely incident radiation. These resonances were identified as either Wood anomalies, where surface plasmons are excited on the bent metal wires of the mesh, or Rayleigh-Wood anomalies, where the amplitudes of the diffraction orders are redistributed at the Rayleigh wavelengths. At oblique incidence, a splitting effect of the Wood anomalies was also observed. It was shown that two adjacent plasmonic resonances can be used to realize a THz passband that is tunable by the incidence angle of the THz radiation. This makes the plain weave an efficient, low-cost, readily available, and mechanically robust THz bandpass filter. Passbands with desired center frequencies below 3 THz are possible using plain weave with different structure sizes. Secondly, the polarization filtering performance of a polymer-metal mesh that combines metallic and polymer wires is demonstrated. A polarization extinction ratio (field) of above 60:1 for frequencies below 3 THz is determined for the electric field direction parallel and perpendicular to the closely-spaced, parallel metal wires of the mesh, which make up the polarization-sensitive elements. The polymer-metal mesh is therefore an efficient THz linear polarizer, which is mechanically robust and suitable for use as large-area component e.g. in THz wireless communications and THz imaging.

## 5. Back matter

**Funding.** European Union's Horizon 2020 research and innovation programme (grant agreement N° 964735 EXTREME-IR); Deutsche Forschungsgemeinschaft (DFG) within the project 468501411–SPP2314 INTEGRATECH under the framework of the priority programme SPP2314 – INTEREST.

**Acknowledgments.** We are grateful to the company GKD – Gebr. Kufferath AG for providing the meshes and to Andres Busch and Markus Knefel for many helpful discussions.

**Disclosures.** The authors declare no conflicts of interest.

**Data availability.** Data underlying the results presented in this paper are not publicly available at this time but may be obtained from the authors upon reasonable request.